\documentclass[conference]{IEEEtran}
\IEEEoverridecommandlockouts

\usepackage{tikz}          
\usepackage{amsmath}
\usepackage{graphicx}
\usepackage{booktabs}
\usepackage{latexsym}   
\usepackage{array}      
\usepackage{multirow}
\usepackage{algorithm}
\usepackage[noend]{algpseudocode}
\usepackage{threeparttable}
\usepackage{paralist}   
\usepackage{xspace}
\usepackage{color}
\usepackage{xcolor}
\usepackage{colortbl}
\usepackage{adjustbox}
\usepackage{balance}
\usepackage{wasysym}
\usepackage{rotating}   
\usepackage{listings}
\usepackage{tabu}
\usepackage{pifont}
\usepackage{framed}
\usepackage[shortlabels]{enumitem}
\usepackage{multirow}
\setenumerate[1]{itemsep=0pt,partopsep=0pt,parsep=\parskip,topsep=2pt}
\setitemize[1]{itemsep=0pt,partopsep=0pt,parsep=\parskip,topsep=2pt}
\setdescription{itemsep=0pt,partopsep=0pt,parsep=\parskip,topsep=2pt}

\usepackage{url}            
\usepackage{hyperref}
\usepackage{cite}

\usepackage[many]{tcolorbox}
\definecolor{promptgreen}{RGB}{238,246,232}
\definecolor{agentblue}{RGB}{222,235,247}
\definecolor{ctfenvorg}{RGB}{255,242,204}
\definecolor{ungergreen}{RGB}{238,246,232}
\definecolor{exploitred}{RGB}{251,229,214}

\newtcolorbox{promptbox}[1]{
    enhanced,
    breakable,
    boxrule=1pt,  %
    fontupper=\small,
    fonttitle=\bfseries\color{black},
    arc=3pt,  %
    rounded corners,
    colframe=black,
    colbacktitle=promptgreen,
    colback=promptgreen,
    title=#1,
    left=2mm,  %
    right=2mm,  %
    top=1mm,  %
    bottom=1mm  %
}

\newtcolorbox{agentbox}{
        colback=agentblue,
        colbacktitle=agentblue,
        arc=5pt,
        fontupper=\small,
        fonttitle=\bfseries\color{black},
        boxrule=0.5mm,
        boxsep=1mm,
        width=\linewidth,
        breakable,
        title={CTFAgent},
        rounded corners,
        toptitle=1mm,
        lower separated=false
}

\newtcolorbox{ctfenvbox}{
        colback=ctfenvorg,
        colbacktitle=ctfenvorg,
        arc=5pt,
        fontupper=\small,
        fonttitle=\bfseries\color{black},
        boxrule=0.5mm,
        boxsep=1mm,
        width=\linewidth,
        breakable,
        title={CTF Environment},
        rounded corners,
        toptitle=1mm,
        lower separated=false
}

\newtcolorbox{underbox}{
        colback=ungergreen,
        colbacktitle=ungergreen,
        arc=5pt,
        fontupper=\small,
        fonttitle=\bfseries\color{black},
        boxrule=0.5mm,
        boxsep=1mm,
        width=\linewidth,
        breakable,
        title={RAG Understanding},
        rounded corners,
        toptitle=1mm,
        lower separated=false
}

\newtcolorbox{exploitbox}{
        colback=exploitred,
        colbacktitle=exploitred,
        arc=5pt,
        fontupper=\small,
        fonttitle=\bfseries\color{black},
        boxrule=0.5mm,
        boxsep=1mm,
        width=\linewidth,
        breakable,
        title={RAG Exploiting},
        rounded corners,
        toptitle=1mm,
        lower separated=false
}

\usepackage{float}

\usepackage{amsthm}
\newtheorem{definition}{Definition}

\usepackage{xpatch}
\makeatletter
\chardef\TPT@@@asteriskcatcode=\catcode`*
\catcode`*=11
\xpatchcmd{\threeparttable}
  {\TPT@hookin{tabular}}
  {\TPT@hookin{tabular}\TPT@hookin{tabu}}
  {}{}
\catcode`*=\TPT@@@asteriskcatcode
\makeatother
\usepackage{tcolorbox}
\tcbuselibrary{breakable, skins}
\tcbset{%
  label begin/.style={label={#1}},
  label end/.style={after upper=\label{#1}}
}
\newtcolorbox[%
auto counter]{mybox}[2][]{%
  enhanced jigsaw,
  breakable,
  #1}

\newcommand{\distance}{2pt}
\newcommand{\blue}[1]{\textcolor[rgb]{0.00,0.00,1.00}{#1}}

\definecolor{wheat1}{rgb}{1.000000,0.905882,0.729412}
\definecolor{LightGray}{rgb}{0.827451,0.827451,0.827451}

\newcolumntype{a}{>{\columncolor{wheat1}}l}

\definecolor{mygreen}{rgb}{0,0.6,0}
\definecolor{mygray}{rgb}{0.5,0.5,0.5}
\definecolor{mymauve}{rgb}{0.58,0,0.82}
\definecolor{darkblue}{rgb}{0.0,0.0,0.6}
\definecolor{maroon}{RGB}{102, 0, 0}
\definecolor{Maroon}{cmyk}{0,0.87,0.68,0.32}
\definecolor{darkred}{RGB}{139, 0, 0}
\definecolor{forestgreen}{RGB}{34, 139, 34}

\lstdefinelanguage{XML}
{
  basicstyle=\ttfamily\small,   
  morestring=[b]",
  moredelim=[s][\color{darkblue}]{<}{\ },
  moredelim=[s][\color{darkblue}]{</}{>},
  moredelim=[l][\color{darkblue}]{/>},
  moredelim=[l][\color{darkblue}]{>},
  morecomment=[s]{<?}{?>},
  morecomment=[s]{<!--}{-->},
  stringstyle=\color{darkred},
  identifierstyle=\color{mymauve}
}

\lstdefinestyle{customJava}{
  breaklines=true,
  keepspaces=true,
  frame=single,
  language=Java,
  showstringspaces=false,
  basicstyle=\footnotesize\ttfamily,
  keywordstyle=\color{blue},
  otherkeywords={+, getIntent},
  numbers=left,
  numbersep=5pt,
  numberstyle=\scriptsize\color{black},
  rulecolor=\color{black},
  stepnumber=1,
  tabsize=2,
  commentstyle=\itshape\color{green!40!black},
  stringstyle=\color{orange},
  emph=[1]  
  {
        do,
        try,
        new,
        catch,
        while,
        SecProvider,
        SecReceiver,
        SecService,
        SecActivity,
        SecSink,
  },
  emphstyle=[1]{\color{darkred}},
  emph=[2]  
  {
        @Override,
  },
  emphstyle=[2]{\color{purple!40!black}},
  belowskip=-1em, 
}

\newif\ifANNOYMIZE
\ANNOYMIZEfalse

\newif\ifACM
\ACMfalse 

\ifACM
\fi

\ifACM
\newcommand{\myfig}{Figure\xspace}
\else
\newcommand{\myfig}{Fig.\xspace}
\fi

\ifACM
\newcommand{\mysec}{\S}
\else
\newcommand{\mysec}{\S}
\fi

\newcounter{findingCounter}

\newcounter{knowledgeCounter}

\definecolor{cadmiumgreen}{rgb}{0.0, 0.42, 0.24}

\newsavebox{\bigimage} 

\usepackage{listings, xcolor}

\definecolor{verylightgray}{rgb}{.97,.97,.97}
\definecolor{codegreen}{rgb}{0,0.55,0}

\lstdefinelanguage{Solidity}{
	keywords=[1]{anonymous, assembly, assert, balance, break, call, callcode, case, catch, class, constant, continue, constructor, contract, debugger, default, delegatecall, delete, do, else, emit, event, experimental, export, external, false, finally, for, function, gas, if, implements, import, in, indexed, instanceof, interface, internal, is, length, library, log0, log1, log2, log3, log4, memory, modifier, new, payable, pragma, private, protected, public, pure, push, require, return, returns, revert, selfdestruct, send, solidity, storage, struct, suicide, super, switch, then, this, throw, transfer, true, try, typeof, using, value, view, while, with, addmod, ecrecover, keccak256, mulmod, ripemd160, sha256, sha3}, 
	keywordstyle=[1]\color{blue}\bfseries,
	keywords=[2]{address, bool, byte, bytes, bytes1, bytes2, bytes3, bytes4, bytes5, bytes6, bytes7, bytes8, bytes9, bytes10, bytes11, bytes12, bytes13, bytes14, bytes15, bytes16, bytes17, bytes18, bytes19, bytes20, bytes21, bytes22, bytes23, bytes24, bytes25, bytes26, bytes27, bytes28, bytes29, bytes30, bytes31, bytes32, enum, int, int8, int16, int24, int32, int40, int48, int56, int64, int72, int80, int88, int96, int104, int112, int120, int128, int136, int144, int152, int160, int168, int176, int184, int192, int200, int208, int216, int224, int232, int240, int248, int256, mapping, string, uint, uint8, uint16, uint24, uint32, uint40, uint48, uint56, uint64, uint72, uint80, uint88, uint96, uint104, uint112, uint120, uint128, uint136, uint144, uint152, uint160, uint168, uint176, uint184, uint192, uint200, uint208, uint216, uint224, uint232, uint240, uint248, uint256, var, void, ether, finney, szabo, wei, days, hours, minutes, seconds, weeks, years},	
	keywordstyle=[2]\color{teal}\bfseries,
	keywords=[3]{block, blockhash, coinbase, difficulty, gaslimit, number, timestamp, msg, data, gas, sender, sig, value, now, tx, gasprice, origin},	
	keywordstyle=[3]\color{violet}\bfseries,
	identifierstyle=\color{black},
	sensitive=false,
	comment=[l]{//},
	morecomment=[s]{/*}{*/},
	commentstyle=\color{codegreen}\ttfamily,
	stringstyle=\color{red}\ttfamily,
	morestring=[b]',
	morestring=[b]"
}

\ifCLASSOPTIONcompsoc
 \usepackage[caption=false,font=normalsize,labelfont=sf,textfont=sf]{subfig}
\else
 \usepackage[caption=false,font=footnotesize]{subfig}
\fi

\def\BibTeX{{\rm B\kern-.05em{\sc i\kern-.025em b}\kern-.08em
    T\kern-.1667em\lower.7ex\hbox{E}\kern-.125emX}}
\begin{document}

\title{Rethinking and Exploring String-Based Malware Family Classification in the Era of LLMs and RAG}

\ifANNOYMIZE
\author{Anonymous Submission}
\else
\author{
\IEEEauthorblockN{Yufan Chen$^1$, Daoyuan Wu$^2$$^*$\thanks{*: Daoyuan Wu is the co-first and corresponding author.}, Juantao Zhong$^2$, Zicheng Zhang$^3$, Debin Gao$^3$, Shuai Wang$^4$,\\Yingjiu Li$^5$, Ning Liu$^{1,6}$, Jiachi Chen$^7$, and Rocky K. C. Chang$^8$}
\IEEEauthorblockA{$^1$ CityU Shenzhen Research Institute, \texttt{yfchen2224@cityu.edu.cn} \\
$^2$ Lingnan University, \texttt{\{daoyuanwu|ericzhong\}@ln.edu.hk}\\
$^3$ Singapore Management University, \texttt{zczhang.2020@phdcs.smu.edu.sg, dbgao@smu.edu.sg}\\
$^4$ The Hong Kong University of Science and Technology, \texttt{shuaiw@cse.ust.hk}\\
$^5$ University of Oregon, \texttt{yingjiul@uoregon.edu}\\
$^6$ City University of Hong Kong, \texttt{ninliu@cityu.edu.hk}\\
$^7$ Sun Yat-sen University, \texttt{chenjch86@mail.sysu.edu.cn}\\
$^8$ Calvin University, \texttt{rocky.chang@calvin.edu}
}
}
\fi

\maketitle
\begin{abstract}

Malware family classification aims to identify the specific family (e.g., \texttt{GuLoader} or \texttt{BitRAT}) a malware sample may belong to, in contrast to malware detection or sample classification, which only predicts a Yes/No outcome.
Accurate family identification can greatly facilitate automated sample labeling and understanding on crowdsourced malware analysis platforms such as VirusTotal and MalwareBazaar, which generate vast amounts of data daily.  
In this paper, we explore and assess the feasibility of using traditional binary string features for family classification in the new era of large language models (LLMs) and Retrieval-Augmented Generation (RAG).
Specifically, we investigate how \textit{Family-Specific String} (FSS) features can be utilized in a manner similar to RAG to facilitate family classification.
To this end, we develop a curated evaluation framework covering 4,347 samples from 67 malware families, extract and analyze over 25 million strings, and conduct detailed ablation studies to assess the impact of different design choices in four major modules, with each providing a relative improvement ranging from 8.1\% to 120\%.

\end{abstract}

\section{Introduction}
\label{sec:intro}

Malware family classification plays a pivotal role in threat intelligence~\cite{schiavoni2014phoenix, wang2020you} and malware analysis pipelines~\cite{VirusTotal, MalwareBazaar}.
Rather than merely determining whether a sample is malicious, family classification aims to assign a specific family label (e.g., \texttt{GuLoader} and \texttt{BitRAT}) that reflects common behavioral traits, infection vectors, and capabilities.
This fine-grained family attribution is essential for tracking malware evolution, generating automated YARA rules~\cite{naik2020evaluating}, and supporting downstream tasks like clustering, triage, and prioritization.
In particular, crowdsourced malware analysis platforms such as VirusTotal~\cite{VirusTotal} and MalwareBazaar~\cite{MalwareBazaar} require accurate family identification to facilitate automated sample labeling and understanding on their vast amounts of data daily.

Prior research has explored various strategies for malware classification, ranging from manual signature engineering~\cite{schultz2001} and behavior-based heuristics~\cite{rieck2008, huang2016} to learning representations from structural graphs~\cite{wilhelm2023}, disassembled code~\cite{dahl2015}, and API call traces~\cite{rieck2008, huang2016, wu2021program}.
However, despite their potential, these approaches often suffer from significant limitations: handcrafted features do not generalize well~\cite{aghakhani2020malware}; dynamic analysis~\cite{yan2012droidscope, yin2007panorama} is expensive and can be evaded~\cite{kirat2015malgene}; and learned representations, particularly those derived from binary content, tend to be opaque, hard to interpret, and vulnerable to adversarial obfuscation~\cite{yang2021cade}.
As a result, most recent studies treat strings merely as auxiliary inputs; see \mysec\ref{sec:background}.

In this paper, we reconsider a long-standing but underutilized resource in malware analysis: string artifacts extracted from binaries.
These strings can contain human-readable indicators such as command-line options, configuration paths, encryption routines, URLs, domain names, registry keys, and API call names. When properly filtered and semantically understood, these features can offer rich insights into malware behavior.
Historically, string features have been overshadowed by more complex modalities such as control flow or dynamic behaviors.
However, we argue that the rise of large language models (LLMs)~\cite{ouyang_training_2022, touvron2023llama}, especially when combined with retrieval-augmented generation (RAG)~\cite{Lewis_Perez_Piktus_Petroni}, presents a new opportunity to revisit string-based malware family classification.

Our work proposes an exploratory pipeline centered on a new feature abstraction: \textit{Family-Specific Strings (FSS)}.
We define FSS as strings that appear within a specific malware family but are absent in other families.
By curating a database of these strings from a labeled training set, embedding them into a semantic vector space, and matching them against features extracted from new samples, we aim to \textit{explore and assess the feasibility of using traditional binary string features for malware family classification in the era of LLMs and RAG}.
To systematically examine this idea, we design a four-stage pipeline and conduct an extensive empirical study evaluating each component both individually and in combination.
Our goal is \textbf{not} to propose a fixed system, \textbf{but rather} to explore the design space of FSS-centered malware family classification through four key research questions.

The first question (RQ1) concerns string extraction: can static-only extraction methods, such as FLOSS~\cite{FLOSS}, recover enough meaningful strings for robust classification, or is dynamic execution (such as sandbox-based memory snapshots) needed to supplement the feature space? Our analysis shows that static extraction alone leaves many semantically rich strings hidden, especially for heavily obfuscated families like \texttt{Stop} or \texttt{Formbook}, and that hybrid approaches can improve classification accuracy by up to 120\% on average.
This highlights the importance of runtime-based visibility for fully capturing string semantics. 

The second question (RQ2) addresses how to filter and organize the training-time string corpus to build a high-quality, semantically meaningful FSS vector database.
Naively embedding all raw strings is computationally infeasible and introduces semantic noise.
To address this, we compare a frequency-based filtering method with an LLM-assisted approach that uses lightweight, cheap GPT-3.5 to estimate string meaningfulness.
Our experiments show that LLM-based filtering improves top-1 classification accuracy by 29\%. 
We further investigate whether intra-family string clustering can improve feature compactness and retrieval quality, laying the groundwork for structurally aware database construction.

The third question (RQ3) considers the test-time phase: given that a malware binary may yield thousands of strings, which subset should be selected as the query input to retrieve matching FSS features?
We compare random subsampling with a k-means clustering-based method that selects centroid strings using TF-IDF representations~\cite{salton1988term}.
The clustering-based method consistently outperforms random sampling, improving top-1, top-2, and top-3 classification accuracy by 11.1\%, 7.1\%, and 9.3\%, respectively, especially for samples with extensive or diverse string content.
This suggests that semantically representative string selection is a critical factor in retrieval-based classification pipelines.

The final question (RQ4) concerns the choice of inference model.
Once the top-k candidate FSS features are retrieved for a test sample, should they be scored using vector similarity (that is, weighted cosine similarity over family-level embeddings), or should an LLM fine-tuned on classification prompts be used to reason over the retrieved features and infer the family label?
Interestingly, our results show that vector-based scoring slightly outperforms the fine-tuned LLM approach (with a relative gain between 8.1\% and 11.9\%), despite the latter's semantic capabilities.
Error analysis reveals that LLMs struggle with noisy or unparseable input strings, which make up a significant portion of many binaries.
Conversely, vector scoring suffers from higher confusion among semantically similar families.
These results suggest complementary strengths, and we identify hybrid inference as a promising direction.

Our experiments are based on a dataset of 4,347 samples from 67 malware families, curated from MalwareBazaar with strict temporal and validation constraints to simulate real-world classification settings.
We extract over 25 million raw strings, apply staged filtering and clustering, and build a vector database indexed by meaningful FSS features.
Our pipeline enables systematic exploration of the impact of the four critical design choices described above, supports detailed ablation studies, and lays a foundation for future research into LLM- and RAG-based malware classification.

\noindent
\textbf{Artifact and Ethics.}
We release the artifact at \url{https://github.com/AIS2Lab/MalwareGPT}.
Experiments were conducted within our own controlled environment. 

%
%
%
%

\section{Preliminaries}
\label{sec:background}

\subsection{Strings in Binary Samples}
\label{sec:background-string}

Strings embedded in binary executables often reveal valuable semantic information, such as API names, configuration parameters, URLs, or command-line instructions. These strings, even when obfuscated, can provide critical clues for identifying malware behavior and origin. In this work, we primarily leverage \textit{FLOSS} (FireEye Labs Obfuscated String Solver)~\cite{FLOSS} to extract various categories of strings from malware binaries. FLOSS is a static analysis tool designed to automatically identify and recover obfuscated strings without requiring dynamic execution. Specifically, it detects the following four types of strings:

\begin{itemize}
    \item \textbf{Static Strings}: These are plain ASCII or UTF-16LE strings stored in readable sections (e.g., \texttt{.rdata}, \texttt{.data}) of the binary. They can be directly extracted without any decoding or emulation. Common examples include hard-coded messages, file paths, or API names.

    \item \textbf{Stack Strings}: These strings are constructed on the stack at runtime, typically through character-wise operations or byte-wise assignments. While not present in the binary as contiguous sequences, stack strings can be recovered through static analysis of instruction patterns that build strings in memory.

    \item \textbf{Tight Strings}: A specific form of stack strings that are tightly encoded and fully decoded within a single function. These are commonly seen in heavily obfuscated malware and are distinguished by their compact encoding and decoding routines.

    \item \textbf{Decoded Strings}: These are strings produced by decryption or decoding routines identified within the binary. FLOSS analyzes known decoding function signatures and instruction sequences to statically recover strings that would otherwise be revealed only at runtime.
\end{itemize}


\subsection{Prior Works Leveraging String Features for Malware Detection and Classification}
\label{sec:background-review}


\begin{table*}[t!]
\centering
\scriptsize 
\caption{Summary of Prior Works Leveraging String Features in Malware Detection and Family Classification.}
\label{tab:related-work-summary}
\setlength{\tabcolsep}{4pt}
\begin{tabular}{p{1.4cm}p{1.1cm}p{1.4cm}p{4.5cm}p{1.2cm}p{7cm}}

\toprule
\textbf{Paper} & \textbf{Venue} & \textbf{Task Type} & \textbf{Feature Type} & \textbf{String Role} & \textbf{How Strings Are Used in the Analysis} \\
\midrule

Schultz et al.\cite{schultz2001} & S\&P\newline(2001)  & Detection & Strings + n-grams + PE header (static features) & Primary & Data mining using string patterns and n-grams \\ 
\hline

Rieck et al. \cite{rieck2008}  & DIMVA\newline(2008) & Detection & API call sequences (dynamic behavior) & Auxiliary & Generate a feature vector for every malware, including the frequency of some specific strings. \\
\hline

Dahl et al. \cite{dahl2015} & MALWARE\newline(2015) & Detection & Printable string histograms\newline+ binary metadata & Auxiliary & String hash values are used in histogram-based classification; Extract the frequency of selected keywords from disassembled code as feature. \\ 
\hline

Siddiqui et al. \cite{siddiqui2016} & CODASPY\newline(2016) & Family Classification & Opcode, entropy, and behavior fusion (static) & Auxiliary &  Only histograms related to the string length distribution are used. \\
\hline

Huang et al. \cite{huang2016} & DIMVA\newline(2016) & Detection\newline+ Family & Dynamic behavior logs (sandbox-based) & Auxiliary & Behavioral signatures are generated by correlating runtime string patterns with system calls. \\ 
\hline

Dambra et al. \cite{dambra2023} & CCS\newline(2023) & Detection\newline+ Family & Static + dynamic features (hybrid) & Auxiliary & Static features outperform dynamic;\newline String features had marginal impact. \\
\hline

Aonzo et al. \cite{aonzo2023} & USENIX\newline(2023) & Family Classification & Human decision features vs. ML features & Auxiliary & Humans and ML shared decision features; strings indirectly referenced. \\
\hline

Wilhelm et al. \cite{wilhelm2023} & RAID\newline(2023) & Detection (evasive) & Behavior summaries (evasion-resilient) & Auxiliary & Behavioral features extracted from string patterns and correlated with API call sequences \\ 
\bottomrule

\end{tabular}
\end{table*}

Malware detection and classification have long relied on extracting and engineering robust features from executable files or their behaviors. Over time, researchers have adopted increasingly diverse features, ranging from static structures like n-grams and headers, to dynamic traces such as API call sequences and behavior logs, as well as domain-specific patterns like DGA-generated URLs. Among these, string features have appeared intermittently, playing different roles depending on task objectives, analysis methods, and modeling choices.

Table~\ref{tab:related-work-summary} provides a chronological overview of nine representative studies published between 2001 and 2024, analyzing their respective goals (malware detection vs. family classification), feature engineering strategies, and, specifically, how string-based features were handled. The table introduces six dimensions: (1) paper metadata, (2) venue and year, (3) task type, (4) feature type, (5) string role, and (6) how strings are used. This comparison reveals both the historical evolution and methodological divergence in the field.

In early work, Schultz et al.~\cite{schultz2001} pioneered the use of string features as core discriminative inputs. By leveraging printable strings and n-gram sequences extracted statically from binaries, their system demonstrated strong detection performance, even outperforming traditional antivirus signatures. This underscores an important historical moment when strings were treated as primary signals.
Subsequent work, such as Rieck et al.~\cite{rieck2008} and Huang et al.~\cite{huang2016}, moved away from static indicators and instead embraced dynamic behavior logs and execution traces. These systems achieved greater generalization and robustness but excluded string features altogether, reflecting a broader trend of relying on runtime behavior as malware authors increasingly obfuscated static content.

Between these works, some studies have used string features in auxiliary roles. Dahl et al.~\cite{dahl2015} included hashed printable strings as part of their input feature histograms, showing that string content can be aggregated but also abstracted away. Dambra et al.~\cite{dambra2023}, in a comprehensive comparison of static and dynamic features, observed that string-based indicators provided marginal but meaningful improvements, highlighting their potential when carefully filtered and combined.

Other studies like Siddiqui et al.~\cite{siddiqui2016} and Wilhelm et al.~\cite{wilhelm2023} explicitly avoided using strings in favor of richer statistical or behavior-based representations. Siddiqui et al. focused on entropy and opcode-level features for family classification, while Wilhelm et al. built classifiers on abstracted behavior summaries, which are useful against evasive samples.
Recent work by Aonzo et al.~\cite{aonzo2023} explored human-versus-machine feature reliance and found that strings, although not explicitly used as input, appeared in the sandbox reports referenced by both humans and models. This suggests that string-level cues continue to inform decision-making, even if only implicitly.



\subsection{Rethinking String-based Malware Classification in the New Era of LLMs and RAG}
\label{sec:background-rethink}

Large Language Models (LLMs), including prominent examples such as the GPT series~\cite{ouyang_training_2022} and LLaMa series~\cite{touvron2023llama,Touvron_Martin_Stone_Albert_Almahairi_Babaei_Bashlykov_Batra_Bhargava_Bhosale_et}, represent a significant advancement in natural language processing. Trained on massive, diverse textual corpora, these models exhibit remarkable performance across tasks such as text generation and question answering.

However, general-purpose LLMs, while powerful, often lack domain-specific precision because their pretraining data is static.
To address this, two complementary strategies have emerged: \textit{fine-tuning}~\cite{ding2023parameter} and \textit{Retrieval-Augmented Generation} (RAG)~\cite{Lewis_Perez_Piktus_Petroni}.
Fine-tuning customizes a pretrained LLM for a particular domain using a smaller, task-specific dataset, improving its ability to produce accurate and context-sensitive responses.
In the software engineering domain, fine-tuned models such as CodeLLaMa~\cite{roziere2023code} have proven effective in code generation~\cite{weyssow2023exploring, li2024extracting}, program repair~\cite{silva2023repairllama, huang2025comprehensive}, vulnerability auditing~\cite{sun2024gptscan, ma2024combining}, and dynamic analysis~\cite{zhong2025defiscope, zhang2025low}.
Meanwhile, RAG enhances model outputs by retrieving relevant textual information from an external corpus at inference time~\cite{Lewis_Perez_Piktus_Petroni}, allowing LLMs to access up-to-date, domain-specific content~\cite{zhang2024acfix, LLM4Vuln, PropertyGPT25, ji2025measuring} without additional training.

These advances motivate a fundamental rethinking of how string-based features, long regarded as static, noisy, or superficial, can be integrated into malware classification pipelines. In traditional systems, string features, such as API names, file paths, registry keys, or command-line options, have often been treated as auxiliary or ignored entirely in favor of behavioral or statistical indicators. Yet, as early research has shown, strings contain rich semantic content and are often highly discriminative, especially when considered at the family level.


With the emergence of LLMs and retrieval-augmented reasoning, we now have models that excel at interpreting, comparing, and contextualizing short, unstructured textual data, which is exactly the form that string artifacts take. LLMs are particularly well suited to leverage the latent semantics in malware strings, even without precise structural features or behavioral logs. In addition, the RAG paradigm offers a natural way to connect an unknown sample’s string footprint with a curated corpus of family-specific knowledge, enabling robust semantic alignment without the need for explicit symbolic rules or costly model fine-tuning.

Rethinking string-based classification under this new trend or paradigm involves (i) curating a high-quality, interpretable set of family-specific strings, (ii) constructing retrieval pipelines that match new samples to relevant historical string contexts, and (iii) using LLMs or vector-based methods to reason about the relationships between these artifacts and known malware semantics.
This approach
offers a novel and interpretable path forward for malware family classification.

\begin{figure*}[t!]
    \centering
    \includegraphics[width=0.95\linewidth, trim=0pt 40pt 0pt 0pt, clip]{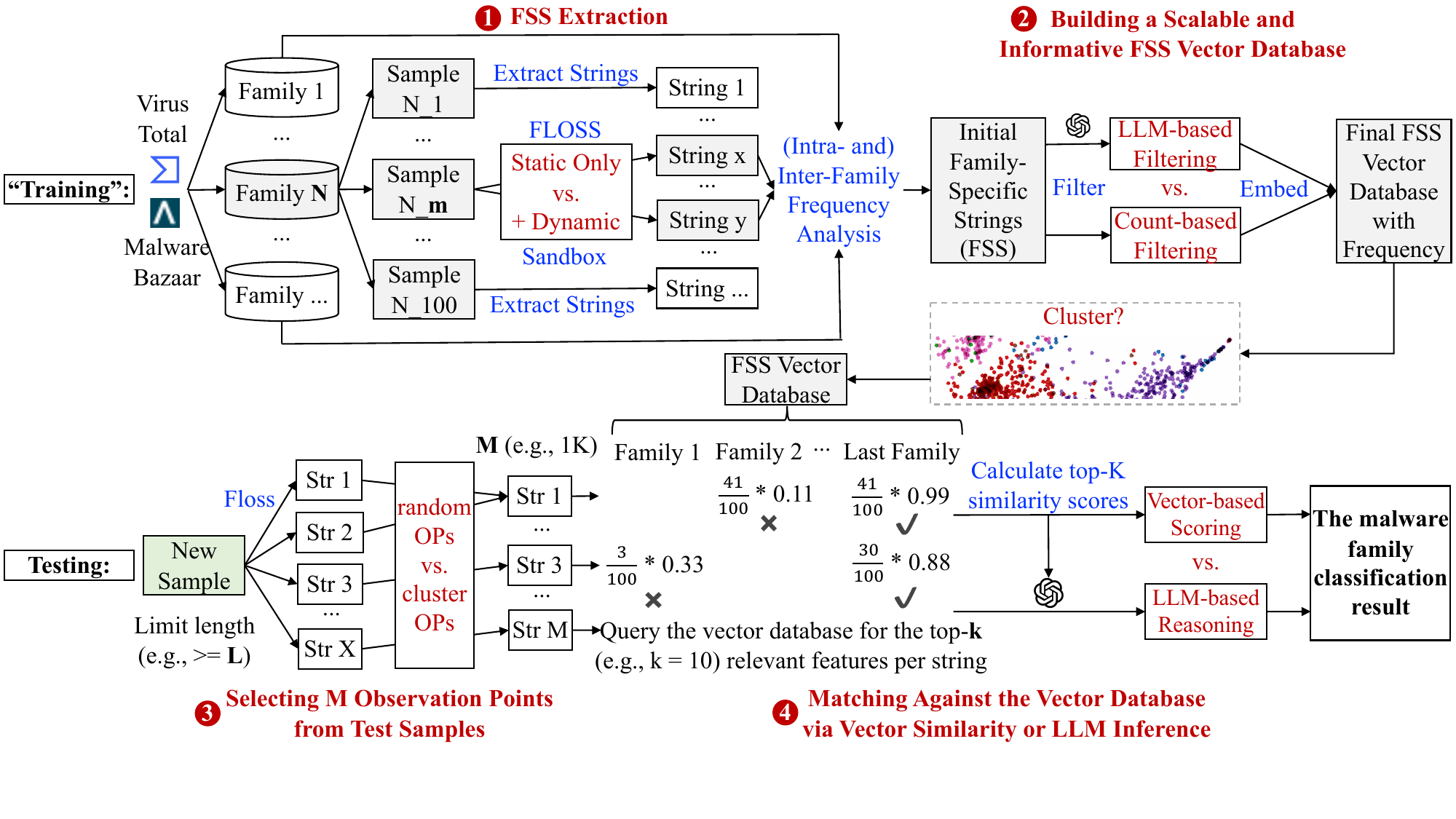}
    \caption{A high-level overview of our exploratory study.}
    \label{fig:overview}
\end{figure*}

\section{An Exploratory Study}
\label{sec:study}

\subsection{Exploration Based on Family-Specific String Features}
\label{sec:overview}

We now turn to a systematic exploration of how such strings can be used in practice. Instead of treating string artifacts as low-level, pre-classification features to be abstracted away, we consider them central to a semantic retrieval and reasoning pipeline. Specifically, we investigate how \emph{Family-Specific Strings (FSS)}, which are interpretable string-level features that are both representative and discriminative of malware families, can be extracted, organized, and utilized throughout the classification process.

We begin by formally defining what constitutes a family-specific string and how it is distinguished from general-purpose string artifacts.
\begin{definition}[Family-Specific String]
\label{def:FSS}
Let $B$ be a malware binary, and let $F$ denote its corresponding malware family. Let $\mathcal{B}$ be the set of all binary samples that belong to family $F$, and let $\mathcal{S}$ be the set of all strings decoded from samples in $\mathcal{B}$. Let $\mathcal{F}$ be the set of all known malware families.

A string $S$ is said to be \emph{family-specific} with respect to $F$ if and only if:
\begin{equation}
    S \in \mathcal{S} \quad \text{and} \quad \not\exists F' \in \mathcal{F} \setminus \{F\} \text{ such that } S \in \mathcal{S}_{F'}
\end{equation}
where $\mathcal{S}_{F'}$ denotes the set of strings decoded from all binaries associated with family $F'$.
\end{definition}
In other words, a family-specific string is one that occurs within the samples of a given family $F$ but does not appear in any samples belonging to other families.

\noindent
\textbf{Overview of the Pipeline.}
\myfig\ref{fig:overview} illustrates the overall pipeline studied in this work. Rather than presenting a fixed system, we decompose the pipeline into four modular components and investigate alternative design choices in each. These components include: (1) string extraction from malware binaries, (2) construction of a searchable vector database of family-level strings, (3) selection of query-time string inputs from test samples, and (4) matching and inference strategies for classification. The goal is to understand how different configurations of these components affect the performance and interpretability of string-centric classification.

This pipeline design supports flexibility in each module, allowing us to isolate and compare key implementation choices. The rest of this section details the research questions and the exploratory experiments conducted in each module.


\subsection{Research Questions}
\label{sec:RQs}

The classification pipeline illustrated in \myfig\ref{fig:overview} comprises four key modules: string extraction, vector database construction, query-time string selection, and matching-based inference. Each module presents distinct design choices that influence the pipeline's overall effectiveness and interpretability. Rather than committing to a single architecture, we treat each module as an independent unit of analysis and explore competing strategies through comparative experiments. Our goal is to surface practical insights into how Family-Specific Strings (FSS) can be leveraged for malware family classification in the era of LLMs and retrieval-augmented inference.
To guide this investigation, we articulate four research questions corresponding to the major design modules:

\textbf{RQ1:} Should family-specific strings be extracted purely from static analysis, or can hybrid approaches that incorporate dynamic execution provide better discriminative features?

\textbf{RQ2:} How should we construct the vector database for family-level string matching?  
(a) How should training strings be compressed: by selecting top strings or using LLM-assisted semantic filtering?  
(b) Should the strings in the database be further clustered to improve matching quality?

\textbf{RQ3:} Given that test-time samples may contain numerous strings, how can we effectively select observation points (OPs) to serve as the query representation?

\textbf{RQ4:} In the final stage of family classification, should the system rely on similarity-based scoring over retrieved strings, or use fine-tuned LLMs for semantic inference?

\subsection{Extracting Family-Specific Strings: Static-Only or Hybrid?}
\label{sec:part1}

The first component of our exploratory pipeline extracts candidate strings from malware binaries to serve as an initial set of raw FSS features.
As illustrated in \myfig\ref{fig:overview},
this component in the offline ``training'' phase includes three steps, introduced below; in particular, in the second step, we conduct an ablation study in \mysec\ref{sec:RQ1} to address a foundational question for FSS‑based classification: should we rely solely on static analysis, or should we also incorporate dynamic execution to recover a broader and potentially more representative set of string features?

\noindent
\ding{172}
\textbf{Data Collection and Pre-Processing.} 
As introduced subsequently in \mysec\ref{sec:setup}, we collect numerous samples for each malware family to represent its variants.
Suppose there are $N$ families, as shown in \myfig\ref{fig:overview}, we collect, process, and obtain $m$ samples for each family, where $m$ could be 100 or 50, depending on the average number of available samples for all families.
Specifically, to prioritize samples with more noticeable FSS features for binary string ``training,'' we filter out samples that are \textit{clearly} packed by performing a packing check using Detect It Easy (DIE)~\cite{DetectItEasy} and an entropy‐based analysis~\cite{lyda2007using}.

\noindent
\ding{173}
\textbf{String Extraction from Binaries.}
Once we have a set of $N \times m$ samples, we need to extract raw strings from binaries.
As mentioned above, there are two design choices: a static-only approach or one that incorporates dynamic execution.
The former is scalable and fully automated, while the latter may not be.
In \mysec\ref{sec:RQ4}, we compare the effect of the following two strategies for different malware families.

\begin{itemize}
\item \textbf{Static-only String Extraction}:
In the static setting, we utilize FLOSS~\cite{FLOSS}, a widely adopted reverse-engineering tool that extracts four categories of printable strings from binaries: \textit{static strings}, \textit{stack strings}, \textit{tight strings}, and \textit{decoded strings}; see a more detailed introduction in \mysec\ref{sec:background-string}.
FLOSS decodes the embedded strings within each sample, producing a set of raw strings for every malware binary in the family.
These strings capture constants, literals, and obfuscated values reconstructed through dataflow analysis.
Static methods offer full coverage and reproducibility without requiring execution, but they may fail to expose runtime-generated strings.

\item \textbf{Incorporating Dynamic String Extraction}:
A malware analysis pipeline can also incorporate dynamic execution, although this comes at the cost of scalability and requires manual effort.
In this paper, we use Falcon SandBox~\cite{FalconSandbox}, a commercially supported, feature-rich malware analysis environment.
This platform provides real-time API call monitoring with comprehensive coverage of Windows system calls, memory forensics with high string-reconstruction accuracy, and multi-path execution tracing for increased branch coverage.
In this hybrid setting, each sample undergoes both FLOSS-based static analysis and Falcon-based dynamic execution.
The resulting sets of strings are then unified, with duplicates removed and invalid entries filtered.
This setup allows us to empirically assess whether augmenting static FSS with dynamic strings improves family classification performance.
\end{itemize}

\noindent
\ding{174}
\textbf{Family Frequency Analysis.}
Based on the definition of FSS in Definition~\ref{def:FSS}, we perform a cross-family comparison of all strings extracted by FLOSS to obtain an initial FSS feature set for each malware family, as illustrated in \myfig\ref{fig:overview}.
In other words, if a string appears in the samples of one family and never appears in any samples of other families, we consider it an FSS.
This step filters out common strings that may appear in multiple families, such as general library calls, common system functions, or widely used file names.
In addition to inter-family frequency analysis, we also analyze the intra-family frequency of FSS features and sort them by frequency, which will be only used for further filtering when constructing the final FSS vector database in \mysec\ref{sec:part2}.


\subsection{Building a Scalable and Informative FSS Vector Database}
\label{sec:part2}

As illustrated in \myfig\ref{fig:overview}, the second component involves constructing a scalable and informative vector database from the initial set of raw FSS features obtained in \mysec\ref{sec:part1}.
It serves as the RAG database that enables query-time retrieval and matching in later stages.
However, the raw FSS feature pool is typically large and noisy. For example, in our training corpus of 3,350 samples from 67 families (\mysec\ref{sec:setup}), the average number of raw FSS strings per family is as high as 334,116 (median: 137,160).
Embedding such a large volume is computationally expensive and semantically redundant.

To address this challenge, a malware analysis pipeline can employ a two-step refinement process:
(1) compressing the raw FSS strings through either LLM-based semantic filtering or top count-based filtering,
and (2) optionally organizing them via vector clustering.
The effectiveness of these design choices is evaluated in ablation studies, as further discussed in \mysec\ref{sec:RQ2}.

\noindent
\textbf{Count-based vs. LLM-based Filtering.}
A straightforward filtering method is to retrain top-K most frequent FSS features.
However, this naive method may ignore FSS features that reflect family-specific features but are not in high count.
To address this problem, we also design a filtering method based on semantic content to refine the feature set.
This alternative design choice is based on the observation that shorter FSS tend to consist of more meaningless, obfuscated strings (e.g., randomized API names), rather than meaningful, human-readable strings.
Leveraging this insight, we employ an LLM-assisted method to determine a length range that primarily captures useful FSS features, thereby enhancing the relevance and interpretability of the embedded features.

\noindent
\textbf{LLM-assisted FSS Semantic Analysis.}
In the alternative LLM-based design choice, we begin by categorizing FSS features based on their string length.
Initial analysis revealed that strings longer than a minimum length are more likely to contain meaningful information, such as identifiable keywords, URLs, or other structured patterns.
Therefore, we perform semantic analysis of FSS features to \textit{indirectly} determine an appropriate length threshold (i.e., $>= L$).
To identify the most valuable FSS features, we employ a prompt-based semantic analysis method, as outlined below:
\begingroup
  \setlength{\parskip}{-2pt}
    \begin{promptbox}{}
        You are an expert in obfuscated string reading, capable of recognizing any meaning from obfuscated strings. Does each of the following strings have any meaning? Answer in a JSON output only (do not provide any other description), where the key is the string number ID and the value is Yes/No.\\
    		1. string1\\
    		2. string2\\
    		...\\
    		500. string500
    \end{promptbox}
\par\endgroup

For each batch of strings (up to 500 at a time), we input them into the GPT-3.5 model's context window.
To mitigate hallucinations and enhance reliability, we use a voting mechanism involving three independent GPT-3.5 agents with different \texttt{temperatures}.
For each string, if at least two agents agree on ``Yes'' (i.e., the string is meaningful), it is considered valuable.
It is worth noting that our LLM-based semantic analysis is used only to infer the length range where FSS features are most valuable.
That said, it does not eliminate any unintelligible strings within the determined length range.

\begin{figure}[t!]
	\centering
	\includegraphics[width=0.48\textwidth, trim=93pt 17pt 91pt 64pt, clip]{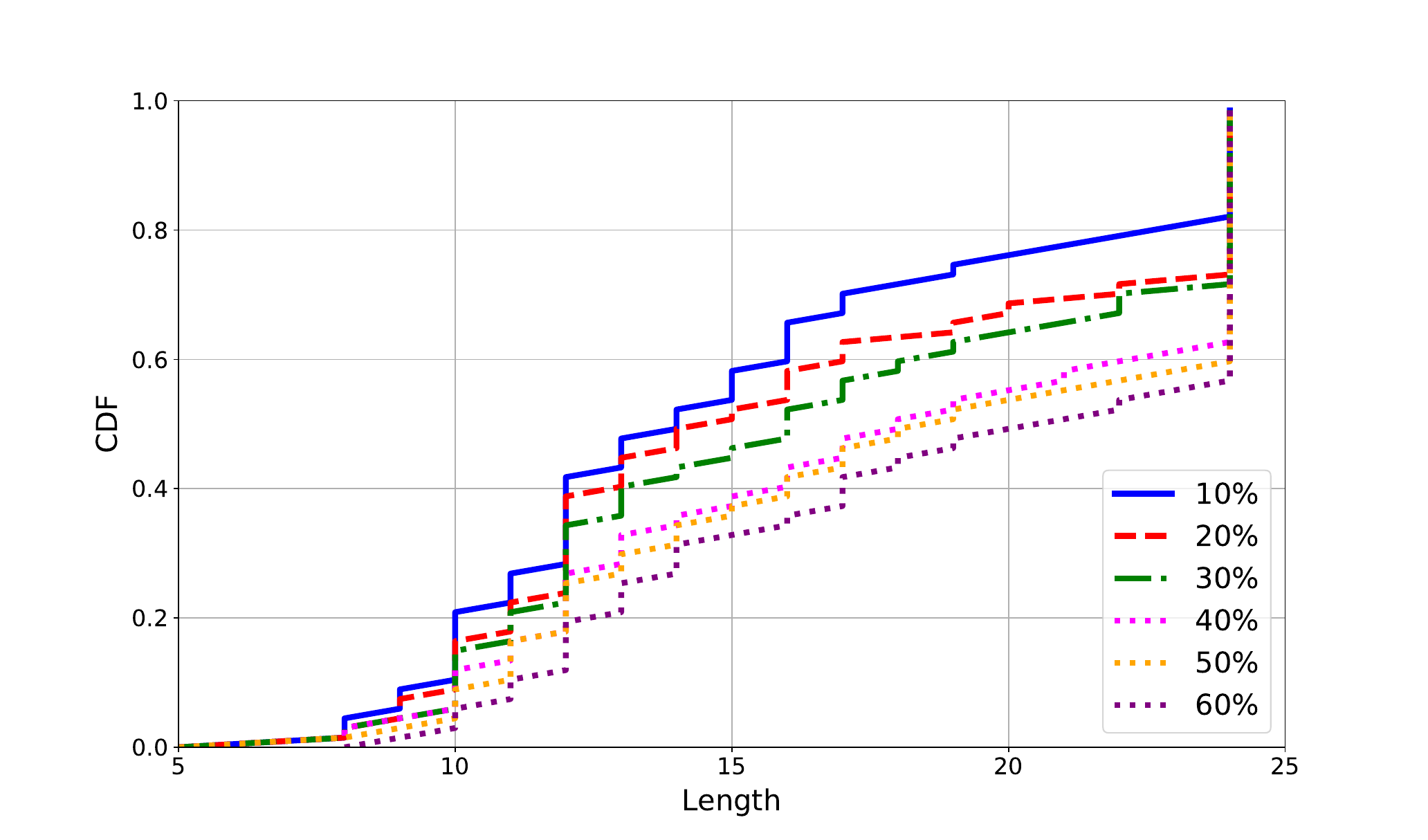}
	\caption{CDF of meaningful FSS at various percentage levels across different FSS lengths for our training set.}
	\label{fig:length_distribution}
\end{figure}

\noindent
\textbf{Determining the Optimal FSS Length Threshold.} 
Based on the LLM analysis results, we calculate the percentage of meaningful FSS features for different lengths.
\myfig\ref{fig:length_distribution} plots the CDF (Cumulative Distribution Function) showing the percentage levels of meaningful FSS across different lengths.
We observe that for our training set in \mysec\ref{sec:setup}, at a length of 13, the minimum percentage of meaningful FSS (i.e., 10\%) now covers over 50\% of all samples.
In other words, for lengths below 13, most samples have a very low percentage of meaningful FSS, often below 10\%, making them less suitable for inclusion in the feature database.
We thus set the length threshold $L$ to 13 for the samples in our training set.
In doing so, we significantly reduce the average (median) number of FSS features per family from 334,116 (137,160) to 33,782 (12,544), achieving an almost 10-fold reduction.

\noindent
\textbf{Embedding the Refined FSS Feature Set into Vector Space.}
With the filtered FSS features through either choice above, we sort them based on the intra-family frequency analysis conducted in \mysec\ref{sec:part1} and retain the top 10,000 strings per family.
This balances the feature set across families and further reduces the size of our vector database.
We then embed this refined set of FSS features into a vector space and store the vector embeddings along with the original plaintext strings, their family labels, and occurrence frequencies. This structure enables efficient approximate nearest-neighbor search, as well as exact match lookup, during test-time matching. 
Note that during embedding the refined FSS features, we also record them separately in plaintext to facilitate fast comparison of common FSS features between a testing sample and the training set, as discussed later in \mysec\ref{sec:part3}. 

\noindent
\textbf{Optional Clustering of Embedded Strings.}  
To further reduce redundancy and improve retrieval interpretability, we also explore clustering the embedded FSS vectors using unsupervised algorithms such as \textit{k}-means~\cite{krishna1999genetic} and DBSCAN~\cite{khan2014dbscan}.
The hypothesis is that grouping semantically similar strings can enhance retrieval stability, reduce query-time noise, and surface core behavioral motifs. We evaluate the impact of clustering in \mysec\ref{sec:RQ2b} through an ablation analysis.

\noindent
\underline{Takeaway:}
Through the above filtering, embedding, optional clustering process, the initial raw FSS features are transformed into a compact and semantically structured knowledge base for test-time retrieval and matching.
The trade-offs between filtering strategies and the effect of clustering are key focus areas in our investigation of FSS-based classification.

\subsection{Selecting Observation Points from Test-Time Samples}
\label{sec:part3}

The third component of our exploratory pipeline concerns how to effectively select \emph{Observation Points (OPs)} from a testing sample to serve as the query representation for family classification.
As shown in \myfig\ref{fig:overview}, each test-time sample typically yields a large number of strings, many of which are noisy, redundant, or irrelevant. Efficiently selecting a meaningful subset of strings is critical for both inference accuracy and computational feasibility.

We explore two competing strategies for selecting OPs: \textit{random subsampling} and \textit{clustering-based selection}.
An ablation study in \mysec\ref{sec:RQ3} quantifies the impact of each strategy on classification accuracy, as motivated by RQ3.

\noindent
\textbf{Baseline Strategy: Random Subsampling.}
In the baseline setting, we randomly sample \( M \) strings (e.g., \( M = 1{,}000 \)) from the FLOSS-extracted string set of the test sample, after applying the length threshold from \mysec\ref{sec:part2}.
Moreover, to ensure discriminative string coverage, any common strings shared between the test sample and the FSS vector database (identified through exact string matching) are always included in the OPs.
While this random subsampling method is fast and simple, it risks omitting semantically important strings when the total string count is high.

\noindent
\textbf{Alternative Strategy: Clustering-based Selection.}
To address the limitations of random sampling, we introduce a clustering-based method for OP selection. The key intuition is that semantically similar strings often form natural groups, and selecting representatives from these groups can preserve the diversity and informativeness of the entire string set.
Specifically, we convert the string set into vector representations using TF-IDF~\cite{salton1988term} weighted 3-gram embeddings,
then apply the \textit{k}-means algorithm with \( k = M \) to cluster the strings.
The string closest to the centroid of each cluster is selected as the OP, ensuring coverage across diverse semantic regions of the string space.

\noindent
\textbf{Comparative Evaluation.} As discussed in \mysec\ref{sec:RQ3}, we evaluate both strategies using a fixed input budget of \( M = 1{,}000 \) strings per sample. On average, the clustering-based approach improves classification accuracy by over 10\% compared to random subsampling, particularly for samples with large and diverse string sets.
While clustering-based OP selection improves performance, it also incurs additional computational overhead during preprocessing.
The trade-off is justifiable in static analysis settings where string extraction is deterministic and preprocessing is performed offline. Future work may explore more efficient embedding schemes or adaptive clustering thresholds based on string entropy or redundancy levels.

\noindent
\underline{Takeaway:}
This module reinforces the need for semantically-aware preprocessing in the test-time pipeline and ensures that test-time representations are both compact and discriminative.


\subsection{Matching Against the Vector Database via Vector Similarity or LLM Inference}
\label{sec:part4}

The final module in our exploratory study concerns how to interpret the relationship between a test-time sample and the family-specific knowledge base constructed from FSS features.
As shown in \myfig\ref{fig:overview}, after retrieving semantically similar features from the vector database, two strategies can be used to perform the final classification:
(1) \textit{vector-based scoring}, which ranks families based on similarity-weighted evidence accumulation, and (2) \textit{LLM-based reasoning}, which uses a fine-tuned large language model to infer the sample
family based on its top-ranked FSS matches.
Our ablation study in \mysec\ref{sec:RQ4} compares these two approaches.

\noindent
\textbf{Vector-based Scoring.}
In this approach, we follow a RAG-like workflow~\cite{lewis2020retrieval, PropertyGPT25} to retrieve and score relevant FSS entries from the database.
For each of the \( M \) selected OPs from the test sample, we query the FSS vector store to retrieve the top-\( k \) (e.g., \( k = 10 \)) semantically similar strings. This results in \( M \times k \) retrieved strings (or ``hints''), each associated with its source family and frequency.
To estimate the relevance of each hint to the test sample, we compute a frequency-normalized similarity score:
\[
\text{Score}_{\text{weighted}} = \frac{F_{\text{feature}}}{N_{\text{family}}} \times \text{Sim}(\text{Str}_{\text{OP}}, \text{Str}_{\text{feature}})
\]
where \( F_{\text{feature}} \) is the frequency of the feature string within its source family, \( N_{\text{family}} \) is the total number of training samples in that family, 
and \( \text{Sim}(\cdot) \) measures semantic similarity in embedding.

All retrieved hints are ranked by their weighted score, and the top-\( K \) matches (e.g., \( K = 100 \)) are used to aggregate family-level scores.
The top-ranked family is selected as the predicted label for the sample.
This method is lightweight.

\begin{figure}[t!]
	\begin{tcolorbox}[title=Prompt for Fine-tuning]
		
		\textbf{System:}\\
		You are an expert in malware analysis, capable of interpreting the sample type and its family (for in-scope malware) based on the top-K feature strings and top-scoring candidate families.
		
		\tcbline
		\textbf{User:}\\
		We have extracted feature strings from a potential malware sample and obtained the top 100 records with the highest scores in an embedding vector database. Below are the most relevant feature strings and their corresponding source families (in brackets) from the given sample (provided as line-by-line input):\\
		\blue{\{List of top-K string (family) mappings\}}\\
		
		Moreover, based on these top feature strings, we have calculated the top potential malware families for the given sample, sorted from highest to lowest likelihood:\\
		\blue{\{List of top-scoring candidate families\}}\\
		
		Based on the above two pieces of information, please determine the sample type (in-scope malware, out-of-scope malware, or benignware) and the top-3 potential families if it is in-scope malware.\\
		
		Please respond in one of the following three formats (do not include any other descriptions):\\
        \blue{The top-3 families are \{family1, family2, family3\}.}
		
		\tcbline
		\textbf{Response:}\\
        The top-3 families are \{top-1 family, family, family\}.
	\end{tcolorbox}
	\caption{Prompt format for fine-tuning-based reasoning over retrieved top-K features and candidate families.}
	\label{fig:finetuningprompt}
    \vspace{2ex}
\end{figure}

\noindent
\textbf{LLM-based Reasoning.}
To enhance semantic interpretation, we further explore an alternative approach where the top-\( K \) retrieved features and top-scoring candidate families are passed to a fine-tuned LLM for final inference.
As shown in \myfig\ref{fig:finetuningprompt}, the model receives two inputs: (1) the top-\( K \) retrieved FSS strings and their associated source families, and (2) a ranked list of candidate families based on vector scores.
The LLM is trained to output 
the top-3 candidate families.
This is formulated as a structured prompt-based classification task using fine-tuning data derived from validation samples processed through the same pipeline.

We prepare training data for fine-tuning by running the first three modules (\mysec\ref{sec:part1}–\mysec\ref{sec:part3}) on a held-out validation set. This ensures the model is exposed to realistic top-\( K \) feature patterns. We fine-tune a cost-effective LLM (e.g., \texttt{gpt-4o-mini}) using this structured dataset, optimizing it to interpret retrieval patterns and make nuanced family-level judgments.


\noindent
\underline{Takeaway:}
This module completes the exploratory pipeline by testing the final classification layer's ability to reconcile raw string evidence with semantic family-level inference, offering a direct response to the challenge posed in RQ4.

\section{Experimental Setup}
\label{sec:setup}

\subsection{Dataset Collection}
\label{sec:dataset}

To support our exploratory study (see \myfig\ref{fig:overview}), we curated a malware dataset from the MalwareBazaar platform~\cite{MalwareBazaar}, which hosts a large corpus of Windows/UNIX malware samples labeled by multiple engines.
Our dataset construction emulates a real-world scenario where analysts must detect newly emerging malware using a historical repository of known families.

We collected 
and curated a subset of 4,347 labeled samples spanning 67 distinct families for training, validation, and testing.
We validated family labels with VirusTotal to remove inconsistently tagged samples.
For each malware family, we selected 60 pre-2023 samples (50 for the database and 10 for fine-tuning) and 5 near-January 2024 samples for testing.
Some families (e.g., DanaBot, Meterpreter, AgentTesla, Metasploit) had fewer than 5 testable samples due to extraction or obfuscation issues.
As summarized in Table~\ref{tab:datasets}, the dataset is divided into:

\begin{itemize}
    \item \textbf{Training Set}: 3,350 samples (50 per family) collected before 2023, used to construct the FSS vector database.
    \item \textbf{Fine-tuning Set}: 670 samples (10 per family), used to fine-tune the LLM for semantic inference (see \mysec\ref{sec:part4}).
    \item \textbf{Testing Set}: 327 samples (up to 5 per family) from early 2024, used to evaluate generalization to recent threats.
\end{itemize}

\begin{table}[t]
\centering
\caption{Summary of datasets used.}
\scalebox{1.0}{
\begin{tabular}{l|r|r}
\hline
\textbf{Dataset} & \textbf{Samples} & \textbf{Families} \\
\hline
Training (Vector DB)       & 3,350 & 67 \\
Fine-tuning Training Set   & 670   & 67 \\
Testing Set (2024 samples) & 327   & 67 \\
\hline
\textbf{Total}             & 4,347 & 67 \\
\hline
\end{tabular}
    }
\label{tab:datasets}
\vspace{2ex}
\end{table}

\subsection{Environment and Parameter Configurations}
\label{sec:env}

Our experiments were conducted on a workstation with 128GB RAM, 32-core CPU, and NVIDIA A100 GPU. The implementation uses Python with \texttt{faiss} for vector indexing and OpenAI APIs for embedding and fine-tuning.

\noindent\textbf{Embedding and Query Pipeline.}
We use the \texttt{text embedding-ada\allowbreak-002} model to convert each FSS string into a 1,536-dimensional embedding vector~\cite{EmbeddingModel}.
During inference, each observation point (OP) from a test-time sample is used to query the FSS vector database.
The system retrieves the top-$k = 10$ semantically similar entries per OP, based on cosine similarity, forming the retrieval foundation for both scoring-based and LLM-based reasoning.

\noindent\textbf{LLM Fine-tuning.}
For semantic inference, we fine-tune the \texttt{gpt-4o\allowbreak-mini-2024-07-18} model using our curated prompt-response pairs. Each prompt encodes the top-$K = 100$ retrieved FSS strings and their associated source families, along with a ranked list of candidate families (see \myfig\ref{fig:finetuningprompt}).
The fine-tuning set of 670 samples is processed using the same pipeline described in \mysec\ref{sec:part1}–\mysec\ref{sec:part3}, ensuring the model learns to interpret retrieved features. 

\noindent\textbf{Default Parameters.}
Unless otherwise specified, we set $M = 1{,}000$ as the number of OP strings selected from each test sample, $k = 10$ as the number of retrieved FSS strings per OP, and $K = 100$ as the total number of top-ranked FSS features used for scoring or LLM-based classification.

\subsection{How Strings Evolve in the Training Set}

To better understand how our exploratory pipeline contributes to refining string-based features, we trace the evolution of string sets throughout the offline training phase.
Specifically, we analyze how the average number of string features per family changes as we successively apply the design choices discussed in \mysec\ref{sec:part1} and \mysec\ref{sec:part2}.
Our experiments were conducted on the 67 malware families used for training.

\noindent
\textbf{Step 1: Raw String Extraction.}
We begin by extracting all printable and decoded strings from each binary using FLOSS, resulting in an initial set of raw strings per sample. On average, this process yields 377{,}034 strings per family, reflecting the large volume and typical redundancy of FLOSS outputs.

\noindent
\textbf{Step 2: Family-Specific Strings.}
Next, we apply inter-family filtering to identify FSS strings that are unique to each family (\mysec\ref{sec:part1}). This step removes strings common across multiple families (e.g., general system API names), reducing the average string count to 334{,}116 per family.

\noindent
\textbf{Step 3: LLM-Based Filtering.}
To eliminate short and often unintelligible strings, we apply the minimum length threshold ($L = 13$), derived via LLM-assisted semantic analysis (\mysec\ref{sec:part2}).
This step filters out a large fraction of noisy strings and reduces the average string count to 33{,}782 per family.

\noindent
\textbf{Step 4: Removing Recursive Strings.}
As an implementation refinement, we remove recursive strings, which are strings composed of repeating self-patterns (e.g., ``ABABABAB'' or ``XYZXYZXYZ'').
These usually result from FLOSS decoding loops or malformed encodings.
This filtering step eliminates another small subset of strings, bringing the average count down slightly to 33,283.

\noindent
\textbf{Step 5: Frequency-Based Compression.}
Finally, to ensure scalability and balance across families, we retain only the top 10,000 FSS features per family based on intra-family frequency (\mysec\ref{sec:part2}). This final pruning step further reduces the average to 7,042 strings per family, a 53-fold reduction from the original FLOSS output, while preserving the most representative and discriminative features.

Through this sequential filtering pipeline, our system transforms a noisy, oversized raw string set into a compact, interpretable, and semantically meaningful feature corpus. This refined dataset forms the backbone of our FSS vector database and directly impacts downstream classification performance.


\begin{table*}[!t]
\footnotesize
\setlength{\tabcolsep}{0pt}   
\centering
\caption{Classification accuracy in different scenarios for samples from various malware families, with RQ1's effect shown in Table~\ref{tab:dyn_accuracy}.}
\label{tbl:overallResultsGrouped}

\newcolumntype{C}{>{\centering\arraybackslash}p{\dimexpr(\linewidth-1.8cm-11\arrayrulewidth)/12\relax}}

\begin{tabular}{p{1.8cm}|CCC|CCC|CCC|CCC}
\toprule
\multirow{2.2}{*}{\textbf{Family}}
& \multicolumn{3}{c|}{\textbf{Normal Classification}}
& \multicolumn{3}{c|}{\textbf{Use Count Filtering--RQ2}}
& \multicolumn{3}{c|}{\textbf{Use Random OPs--RQ3}}
& \multicolumn{3}{c }{\textbf{Use LLM Reasoning--RQ4}} \\*
\cmidrule(lr){2-4}\cmidrule(lr){5-7}\cmidrule(lr){8-10}\cmidrule(lr){11-13}
& \textbf{top1} & \textbf{top2} & \textbf{top3}
& \textbf{top1} & \textbf{top2} & \textbf{top3}
& \textbf{top1} & \textbf{top2} & \textbf{top3}
& \textbf{top1} & \textbf{top2} & \textbf{top3} \\
\midrule
\multirow{2}{*}{\textbf{Average}}
& / & / & / & 0.31 & 0.36 & 0.40 & 0.36 & 0.42 & 0.43 & 0.37 & 0.40 & 0.42 \\
& 0.40 & 0.45 & 0.47 & \textcolor{green!70!black}{+29\%} & \textcolor{green!70!black}{+25\%} & \textcolor{green!70!black}{+17.5\%} & \textcolor{green!70!black}{+11.1\%} & \textcolor{green!70!black}{+7.1\%} & \textcolor{green!70!black}{+9.3\%} & \textcolor{green!70!black}{+8.1\%} & \textcolor{green!70!black}{+12.5\%} & \textcolor{green!70!black}{+11.9\%} \\

\bottomrule
\end{tabular}
\end{table*}

\section{Results and Analysis}
\label{sec:result}


We now evaluate the classification performance of our exploratory pipeline across 67 malware families using the curated testing dataset introduced in \mysec\ref{sec:setup}. Unless otherwise specified, all reported results follow the \textit{Normal Classification} configuration, which incorporates the following default design choices:
\begin{itemize}[left=0pt]
    \item Static-only string extraction using FLOSS (i.e., dynamic analysis is excluded) in RQ1.
    \item LLM-based filtering of FSS without additional clustering of embedded vectors in RQ2.
    \item Clustering-based selection of observation points in RQ3.
    \item Vector-based scoring for family prediction in RQ4.
\end{itemize}

This configuration represents a fully automated pipeline optimized for speed and scalability.
Two modules, dynamic string extraction (RQ1) and vector clustering (RQ2), are omitted in the default pipeline because they require manual analysis or complex orchestration. These components are revisited in later subsections (i.e., \mysec\ref{sec:RQ1} and \mysec\ref{sec:RQ2b}) through ablation studies to assess their potential impact.

Table~\ref{tbl:overallResultsGrouped} summarizes the per-family and average performance under various configurations.
In the default setting (Normal Classification), our pipeline achieves an average top-1 accuracy of 40\% across all test samples, indicating that in 40\% of cases, the correct malware family is ranked first.
When expanding the evaluation window to include the top-2 and top-3 ranked predictions, the average accuracy increases to 45\% and 47\%, respectively.
This incremental improvement (+5\%, +2\%) suggests that the correct family could be ranked among the top candidates, even if not in top-1, underscoring the semantic value of the retrieved string features.

To understand the effect of each module in the pipeline, we organize the subsequent subsections around the four RQs defined in \mysec\ref{sec:RQs}. The presentation proceeds in the following order: starting with \mysec\ref{sec:RQ2} (RQ2), which examines how to build an effective FSS vector database; followed by \mysec\ref{sec:RQ3} (RQ3), which compares different strategies for selecting observation points; then \mysec\ref{sec:RQ4} (RQ4), which contrasts vector-based scoring with LLM-based reasoning for final classification; and finally, \mysec\ref{sec:RQ1} (RQ1), which investigates the effect of incorporating dynamic execution into the string extraction phase. 
Together, these analyses shed light on the strengths and limitations of the different modules in our pipeline, offering insights into how string-based classification can be enhanced through modular design choices in the era of LLMs and RAG.

\subsection{RQ2: How to Build the FSS Vector Database?}
\label{sec:RQ2}

\subsubsection{RQ2.1: LLM-based Filtering vs. Count-based Filtering}
\label{sec:RQ2a}

We evaluate the effectiveness of LLM-based length filtering by conducting comparative experiments across 67 malware families.
Two experimental conditions are configured:

\begin{itemize}
	\item \textbf{Top Count-based Filtering (Baseline)}:
        Retain the top 10,000 most frequent strings per family, without applying any semantic or length-based filtering.
        This simple heuristic yields a top-1 classification accuracy of 31\%, as reported in Table~\ref{tbl:overallResultsGrouped}.

    \item \textbf{LLM-based Length Filtering (Default)}:
        Apply a length threshold (e.g., $L \geq 13$) derived from the LLM-driven analysis described in \mysec\ref{sec:part2}, which aims to exclude semantically uninformative short strings.
        This approach improves top-1 accuracy to 40.3\%, i.e., 0.40 in Table~\ref{tbl:overallResultsGrouped}.
\end{itemize}

As shown in Table~\ref{tbl:overallResultsGrouped}, LLM-based filtering leads to a 29\% relative improvement in top-1 accuracy (from 31\% to 40\%), and similarly improves the top-2 (from 36\% to 45\%) and top-3 (from 40\% to 47\%) metrics by 25\% and 17.5\%, respectively.
These results validate the hypothesis that short or obfuscated strings introduce semantic noise detrimental to classification.

\begin{figure*}[t]
	\centering
	\subfloat[Histogram showing the distribution of similarity delta values for different malware families after LLM-based filtering.]{
		\includegraphics[width=0.47\linewidth, trim=5pt 5pt 5pt 0pt, clip]{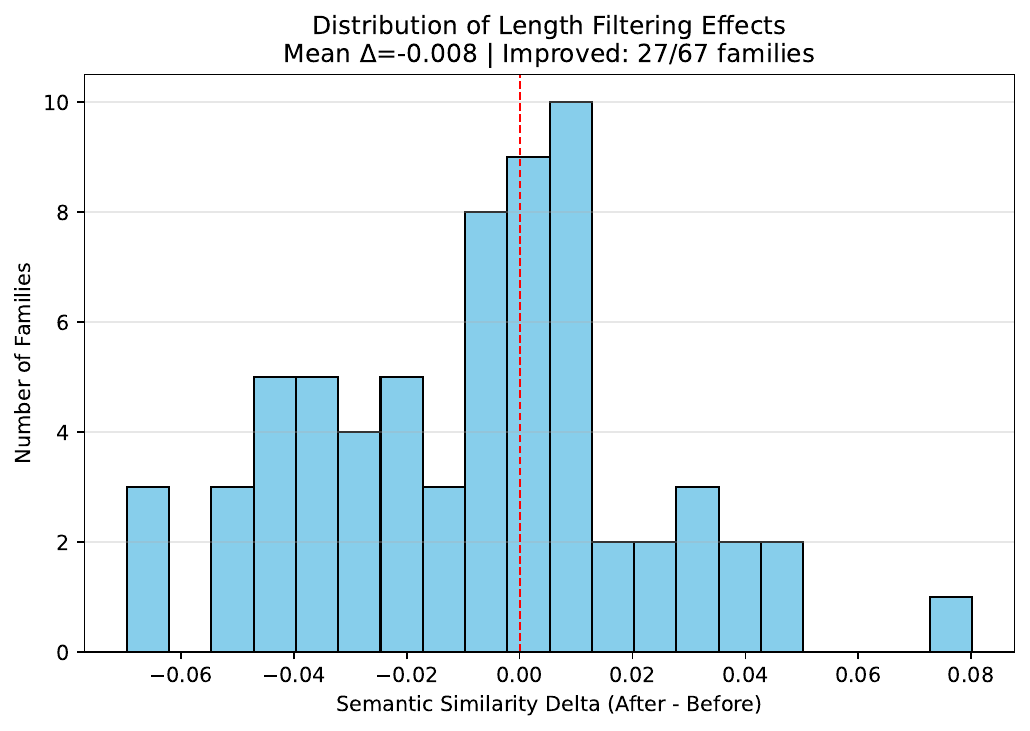}
        \label{fig:delta_dist}
	}
	\hspace{0.02\linewidth}
	\subfloat[Bar chart comparing similarity scores before and after LLM-based filtering for the \texttt{DCRat} family.]{
		\includegraphics[width=0.47\linewidth, trim=40pt 35pt 15pt 42pt, clip]{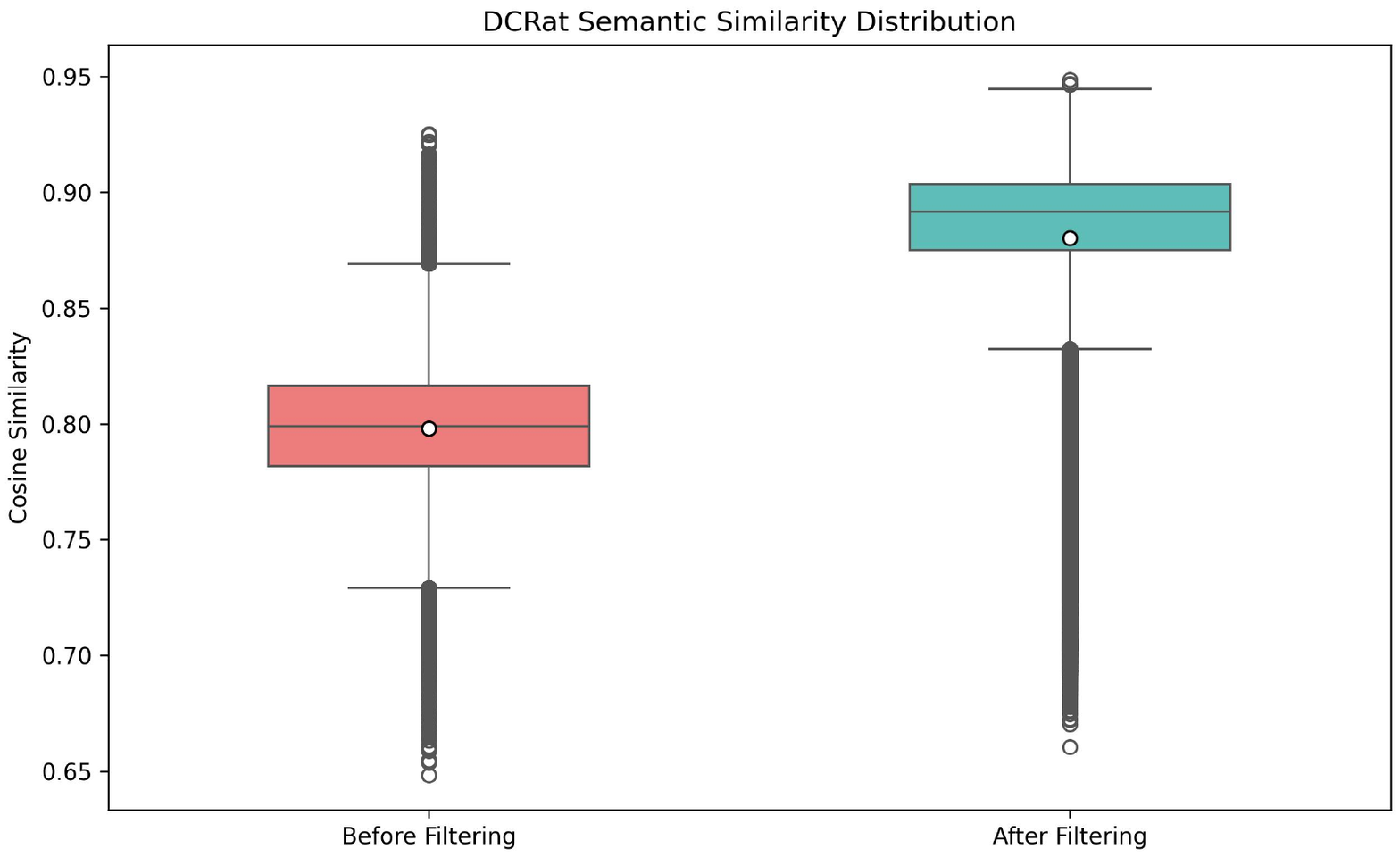}
        \label{fig:dcrat_case}
	}
	\caption{Impact of LLM-based filtering on semantic similarity.}
	\label{fig:rq1_results}
\end{figure*}

To further understand the semantic improvement brought by LLM-based filtering, we analyze changes in intra-family string similarity.
As shown in \myfig\ref{fig:delta_dist}, 40\% of families (27 out of 67) exhibit improved semantic coherence, reflected by positive $\Delta$ values.
The mean improvement across these families is $+0.021$, suggesting that LLM filtering tends to remove noisy or low-information strings that dilute family-specific patterns.

Among the families, \texttt{DCRat} demonstrates the most pronounced benefit, with a delta of $+0.080$ after filtering, as illustrated in \myfig\ref{fig:dcrat_case}.
While the overall improvements are encouraging, the effectiveness of LLM-based filtering is not uniform across all malware families.
We observe a substantial degree of variance in delta values, with a standard deviation of $\sigma_{\Delta} = 0.030$, indicating that \textit{the semantic effect of filtering is highly dependent on malware family characteristics}.

\begin{table}[t]
  \centering
  \caption{Family-wise Changes with LLM-based Filtering.}
  \label{tab:family_perf}
    \scalebox{0.95}{
  \begin{tabular}{lrrr}
    \toprule
    \textbf{Family} & \textbf{Before} & \textbf{After} & \textbf{$\Delta$} \\
    \midrule
    DCRat & 0.720 & 0.800 & +0.080 \\
    CoinMiner & 0.734 & 0.782 & +0.048 \\
    RecordBreaker & 0.792 & 0.723 & $-$0.069 \\
    CobaltStrike & 0.785 & 0.723 & $-$0.062 \\
    \bottomrule
  \end{tabular}
  }
    \vspace{2ex}
\end{table}

As shown in both Table~\ref{tab:family_perf} and \myfig\ref{fig:rq1_results}, the gain in semantic clarity varies significantly depending on the structural characteristics of each family.
We find that filtering is more effective for malware families whose semantic content tends to correlate with string length.
These include families such as \texttt{DCRat} and \texttt{CoinMiner}, where command-and-control (C2) payloads or embedded commands often follow fixed-length patterns.
In contrast, for families such as \texttt{RecordBreaker} and \texttt{CobaltStrike}, where obfuscation is more common and semantically meaningful strings may be short or deliberately fragmented, length-based filtering may inadvertently remove useful features.

These results support our hypothesis that semantic filtering, guided by LLMs, is especially beneficial under two key conditions.
First, when malware families exhibit consistent length-encoded behavior, such as well-formed C2 commands or templated configuration strings, filtering out short, low-entropy strings helps retain high-value patterns.
Second, when the length threshold (e.g., $L = 13$) matches the operational norms of that family, the filtering process achieves a good balance between reducing noise and preserving information.

Nonetheless, the fixed nature of our thresholding strategy introduces limitations.
Our current pipeline adopts a global threshold of $L = 13$ for all families, based on the aggregated LLM analysis results.
However, the family-level heterogeneity observed suggests that this one-size-fits-all approach may be suboptimal.
Future work should explore adaptive or family-specific length thresholds that can better accommodate the diversity in string structures and improve the generalization of semantic filtering across different malware types.

\subsubsection{RQ2.2: How Do Clusters Affect Prediction?}
\label{sec:RQ2b}

Our experiments reveal a significant performance gap across malware families, with about 40\% achieving high classification accuracy (above 0.8), while the remaining 60\% show much weaker results. Upon closer inspection, this variation is strongly associated with the presence or absence of well-formed file clusters, which are subsets of samples within a family that share common frequent strings. These clusters emerge naturally through string co-occurrence patterns. For example, if Sample A contains \{String X, String Y\} and Sample B contains \{String X, String Z\}, both samples are grouped together because they share String X. This clustering behavior suggests structural or behavioral similarities within certain variants of a family.

To better characterize this phenomenon, we group samples into two categories.
First, \textit{Clustered Families} (61.8\% of cases), which exhibit overlapping string patterns and yield an average classification accuracy of 0.87.
Second, \textit{Non-Clustered Samples} (38.2\%), which lack discernible intra-family string consistency and see their accuracy drop to 0.31.
Table~\ref{tab:cluster_performance} summarizes these differences in Jaccard similarity, accuracy, and dataset prevalence.

\begin{table}[t]
	\centering
	\caption{Cluster Types vs. Classification Performance.}
	\label{tab:cluster_performance}
    \scalebox{0.95}{
	\begin{tabular}{lccc}
		\toprule
		\textbf{Cluster Type} & \textbf{Jaccard Similarity} & \textbf{Accuracy} & \textbf{Prevalence} \\
		\midrule
		Clustered & \textgreater0.3 & 0.87 $\pm$ 0.05 & 61.8\% \\
		Non-Clustered & 0 & 0.31 $\pm$ 0.12 & 38.2\% \\
		\bottomrule
	\end{tabular}
    }
    \vspace{1ex}
\end{table}

To understand why some samples do not form clusters, we manually analyze 50 randomly selected non-clustered samples.
Two root causes emerge.
The majority (64\%) are affected by \textit{obfuscation artifacts}, including runtime packers, encryption, and custom string encodings that mask meaningful features.
The remaining 36\% are identified as \textit{genetic variants}, which are samples with substantially different codebases despite sharing the same family label.
These outliers may reflect version changes or parallel development tracks within the same malware lineage.

These findings carry several implications.
First, high performance among clustered families reinforces the value of string-based features when variants retain shared functional components.
Second, the difficulty of classifying non-clustered samples illustrates fundamental limits of static string analysis under heavy obfuscation.
Third, cluster detection itself could be leveraged as a preprocessing heuristic to triage samples: well-clustered inputs can proceed via lightweight string-matching inference, while non-clustered ones can be routed to more advanced methods such as dynamic behavior profiling or graph-based structural comparison.

These observations suggest a need for cluster-aware classification. 
By dynamically identifying the clustering structure of test-time samples, the pipeline can selectively deploy efficient, string-based techniques for homogenous groups and fallback strategies for obfuscated or divergent variants.

\subsection{RQ3: Random Subsampling vs. Clustering-based Selection} 
\label{sec:RQ3} 

In this RQ, we investigate the effect of different OP selection strategies on classification performance.
Since each malware sample may yield thousands of decoded strings, even after filtering, it is critical to select a representative and informative subset for querying the FSS vector database.
To address this, we compare a baseline random sampling method with a clustering-based selection strategy.

\begin{itemize}
	\item \textbf{Random Subsampling (Baseline)}:
        Sample 1,000 strings uniformly at random from each test sample's post-filtered string set.
        While simple and computationally efficient, this strategy often fails to capture semantically diverse strings, particularly in large string sets.

    \item \textbf{Clustering-based Selection (Default)}:
        Perform k-means clustering over the sample's string set using TF-IDF features with character 3-grams and cosine similarity.
        From each cluster, the string nearest to the centroid is selected.
        This ensures semantic diversity by covering multiple regions of the string space, while maintaining a consistent OP count of $M = 1000$.
\end{itemize}


As shown in Table~\ref{tbl:overallResultsGrouped} (RQ3), clustering-based OP selection achieves a top-1 accuracy of 40\%, compared to 36\% using random sampling, representing an 11.1\% relative improvement.
The advantage is especially clear for samples with more than 10,000 decoded strings, where redundancy and noise are more common.
Centroid-based selection improves the representativeness of the OPs, which leads to better alignment with family-specific features in the vector database.

\subsection{RQ4: Vector-based Scoring vs. LLM-based Reasoning} 
\label{sec:RQ4}

In this RQ, we investigate the effectiveness of two competing strategies for malware family classification:
(1) a LLM-based reasoning method (baseline) that employs a fine-tuned LLM to infer malware families from retrieved evidence,
and (2) a vector-based similarity scoring method (default) relying on vector similarity and frequency-weighted scoring.

Table~\ref{tbl:overallResultsGrouped} (RQ4) presents the classification performance for both approaches.
While the vector-based scoring method outperforms LLM-based reasoning at the top-1 level with an 8.1\% relative improvement, its advantage is even more pronounced at the top-2 and top-3 levels.
Specifically, vector-based scoring achieves relative gains of 12.5\% and 11.9\% for top-2 and top-3 classification, respectively.
These results indicate that the vector-based approach is particularly effective at improving broader candidate recall.

To understand the source of performance differences, we conducted a detailed error analysis, as shown in Table~\ref{tab:error_analysis}.
The most common source of error in both approaches is unparseable input, meaning strings that are either nonsensical or the result of obfuscation.
These strings account for 37\% of misclassifications in vector-based scoring and 55\% in LLM-based reasoning.
On the other hand, vector-based scoring is more likely to confuse samples from semantically similar families (25\% compared to 7\% for LLMs), possibly because it relies on surface-level pattern matching without contextual understanding.

These results suggest that LLMs exhibit better semantic generalization in distinguishing closely related families, but are more sensitive to noisy or malformed inputs.
This aligns with expectations: LLM-based reasoning benefits from contextual understanding when inputs are meaningful, but suffers when input quality is low.
In contrast, vector-based scoring, although less robust semantically, handles noisy or superficial patterns more effectively because of its statistical approach.


\begin{table}[t]
    \vspace{-2ex}
	\centering
	\caption{Error Type Distribution Across Approaches.}
	\label{tab:error_analysis}
    \scalebox{0.95}{
	\begin{tabular}{lcc}
		\toprule
		\textbf{Error Type} & \textbf{Vector-based} & \textbf{LLM-based} \\
		\midrule
		Unparseable Strings & 37\% & 55\% \\
		Similar Family Confusion & 25\% & 7\% \\
		\bottomrule
	\end{tabular}
    }
    \vspace{1ex}
\end{table}

Given these observations, we draw three key conclusions:
\begin{enumerate}
	\item In environments where input strings are heavily obfuscated or contain low semantic content, vector-based scoring remains competitive and more stable than LLMs.
	\item LLMs may offer benefits in handling ambiguous or borderline cases, particularly when the distinction between malware families is subtle and pattern-based similarity is insufficient.
	\item A promising future direction lies in hybrid models that combine both strategies, such as using vector scoring to filter out irrelevant candidates and applying LLM-based inference only to the top-ranked clusters.
\end{enumerate}


\subsection{RQ1: Static-only vs. Hybrid String Extraction}
\label{sec:RQ1}

This RQ examines the extent to which dynamic execution improves the quality of extracted strings and, in turn, malware family classification performance.
Our default pipeline uses FLOSS for static-only string extraction due to its scalability and automation.
However, as malware increasingly adopts obfuscation techniques  and runtime packing, static methods alone may fail to uncover semantically useful information.

Our initial, preliminary evaluation revealed that around one third of samples exhibited poor classification accuracy when relying solely on static extraction.
To investigate this, we manually analyze the string artifacts and observe that, even after entropy filtering and regular expression pruning, many static strings remained obfuscated or meaningless.
Table~\ref{tab:string_comparison} presents representative examples contrasting meaningful strings (e.g., API names or URLs) with obfuscated strings (e.g., base64 junk or byte repetitions).
On average, only 31.2\% of extracted strings are semantically interpretable, while 68.8\% are either garbled or machine-generated filler.

\begin{table}[t]
	\centering
    \caption{Comparison of meaningful vs. obfuscated strings.} 
    \scalebox{0.95}{
        \begin{tabular}{p{5cm}p{3cm}} 
		\toprule
		\textbf{Meaningful Strings} & \textbf{Obfuscated Strings} \\
		\midrule
		\texttt{CryptEncrypt}, \texttt{Microsoft\textbackslash Proof\textbackslash hyph32.dll} & \texttt{Xj3\$kP*9@m}, \texttt{aGVsbG8g8J+Yig==} \\
		\midrule
		Average Rate: 31.2\% & Average Rate: 68.8\% \\
		\bottomrule
	\end{tabular}
    }
	\label{tab:string_comparison}
\end{table}

As a comparative experiment, we introduce a hybrid string extraction method that supplements static outputs with dynamic traces collected from Falcon Sandbox.
Falcon provides runtime visibility of system calls, decrypted payloads, and memory-resident strings, many of which are inaccessible through static analysis.
We select five representative families with poor static-only classification results (Loki, Stop, NanoCore, Formbook, and DarkGate) and rerun the pipeline for their 25 samples with dynamic augmentation.

As shown in Table~\ref{tab:dyn_accuracy}, the results are notable.
We observe a 17.2\% increase in the number of semantically valid strings, reflecting Falcon's ability to capture runtime-generated content such as decrypted payloads and system interactions.
Additionally, the proportion of garbage or obfuscated strings drops by 93.4\%, indicating that dynamic execution is effective at filtering out static noise caused by packing and encoding.
The identification of system-level API calls also improves by a factor of 1.1, greatly enhancing the visibility of behavioral indicators that are crucial for malware family classification.
These enhancements lead to clear gains in classification performance, with an average relative improvement of 120\%.
For example, the \texttt{Stop} ransomware family sees its accuracy increase from 0.2 (static-only) to 1.0 (hybrid), due to better visibility into encryption routines, file system interactions, and C2 URLs that are hidden from static inspection.

\begin{table}[t]
	\centering
	\caption{Classification accuracy before and after dynamic execution for 25 samples from five malware families.}
	\label{tab:dyn_accuracy}
    \scalebox{0.95}{
	\begin{tabular}{lccc}
		\toprule
		& \textbf{Static-only} & \textbf{Hybrid} & \textbf{Improvement}\\
		\# Valid Strings & 2,868 & 3,362 & \textcolor{green!70!black}{+17.2\%} \\
        \# Obfuscated Strings & 82,275 & 5,349\textcolor{red!70!black}{(fewer: better)} & \textcolor{green!70!black}{+93.4\%} \\
		\# API Call Strings& 1,220 & 2,558 & \textcolor{green!70!black}{+109.6\%} \\
		\midrule
		\textbf{Average Accuracy} & 0.20 & 0.44 & \textcolor{green!70!black}{+120\%} \\
		\bottomrule
	\end{tabular}
    }
\end{table}

These results confirm the hypothesis that dynamic string extraction offers value for certain malware families. 
While not universally beneficial, hybrid string extraction could be considered a critical capability. 
In future work, we plan to explore partial automation of dynamic execution (e.g., using unpacking stubs or emulated environments) to expand hybrid coverage without sacrificing scalability.

\section{Discussion}
\label{sec:discuss}

Our study reveals both the promise and limitations of string-based malware classification within the context of LLMs and retrieval-augmented pipelines.  
One further extension is the importance of \textit{dynamic length filtering}, tailored specifically to family characteristics.  
While we adopted a fixed threshold ($L=13$) based on aggregated LLM feedback (\mysec\ref{sec:RQ2a}), our results show that families such as \texttt{DCRat} and \texttt{CoinMiner} benefit from this filtering, whereas others like \texttt{CobaltStrike} and \texttt{RecordBreaker} are negatively impacted.  
This highlights the need for adaptive, family-aware filtering to preserve meaningful short strings that might otherwise be discarded.

We also identify limitations in LLM fine-tuning on abstract string data.
Although LLMs excel at code and text-related tasks, their effectiveness in this context is limited (\mysec\ref{sec:RQ4}) for two main reasons: (1) the retrieved FSS strings often lack clear semantics, especially in packed or obfuscated binaries, making it difficult for LLMs to establish reliable associations; and (2) obfuscation artifacts in the training data introduce misleading patterns, which reduces generalization.
These findings suggest that combining RAG-based retrieval with more structured semantic filtering may offer a more robust alternative.

\section{Related Work}
\label{sec:related}

Beyond the summary of string-based classification methods reviewed in \mysec\ref{sec:background-review}, our work is positioned at the intersection of malware analysis and the emerging use of LLMs in cybersecurity.
LLMs have recently shown strong capabilities across a variety of security tasks. For example, \textsc{PentestGPT}~\cite{deng2023pentestgpt} and CHATAFL~\cite{Meng_Mirchev_Bohme_Roychoudhury} have applied LLMs to penetration testing and protocol fuzzing. Other systems have explored fuzzing libraries~\cite{Xia_Paltenghi_Tian_Pradel_Zhang_2024, Deng_Xia_Peng_Yang_Zhang_2023}, program repair~\cite{zhao2024enhancing, zhang2024acfix}, and binary analysis~\cite{Liu_Sun_Zheng_Feng_Qin_Wang_Li_Sun_2023, wong2025decllm}.
Code-focused applications such as LLift~\cite{li2024enhancing} and GPTScan~\cite{sun2024gptscan} further highlight LLMs' strengths in static code understanding.
However, most existing research has overlooked the integration of LLMs with string-level artifacts in malware binaries, an area traditionally considered noisy.
Our work addresses this gap by demonstrating that, with appropriate filtering and semantic grounding, string features can serve as an effective interface between binary artifacts and LLM-based reasoning, providing a lightweight and interpretable alternative to deeper instrumentation or opaque embeddings.

\section{Conclusion}
\label{sec:conclude}

This study revisited string-based malware family classification in the modern context of LLMs and retrieval-augmented architectures.
By introducing the concept of Family-Specific Strings (FSS) and designing a modular pipeline centered on their extraction, organization, and inference, we demonstrated that string artifacts can, when properly curated, yield semantically meaningful and interpretable signals for malware classification. 
Our exploratory analysis across four key design stages revealed that hybrid string extraction improves feature coverage for obfuscated malware, LLM-based filtering enhances semantic consistency, clustering-based observation point selection improves representation quality, and vector-based scoring offers robust and efficient inference.


\section*{AI-Generated Content Acknowledgement}
We used ChatGPT (GPT-4.1) to polish the writing of this paper draft, mainly with prompts such as ``Fix English only.''

\bibliographystyle{IEEEtran}
\bibliography{main}

\end{document}